\documentclass{easychair}
\usepackage{doc}
\usepackage{caption}
\usepackage{subcaption}
\usepackage{booktabs, multirow} 
\usepackage{soul}
\usepackage{changepage,threeparttable} 

\usepackage{listings}

\lstdefinelanguage{coq}{
    keywords={Repair, Lift, module, Module, Theorem, Proof, Record, Lemma, Definition, Abort, Qed, forall, Inductive, Type, Prop, Set, fun, fix, forall, Require, Import, Fixpoint, match, end, with, as, return, struct, Qed, Defined, let, Parameter, Axiom, Patch, Configure, Preprocess},
    basicstyle=\linespread{0.95}\small\ttfamily,
    keywordstyle=\color{blue},
    commentstyle=\itshape\rmfamily,
    showstringspaces=false,
    columns=flexible,
    breaklines=true,
    texcl=true,
    mathescape=true,
    tabsize=4,
    stringstyle=\color{brown},
    escapeinside={(@}{@)},
}

\title{Getting More out of Large Language Models for Proofs}

\author{
Shizhuo Dylan Zhang\inst{1}
\and
   Emily First\inst{2}
\and
    Talia Ringer\inst{1}
}

\institute{
  University of Illinois Urbana-Champaign, USA
\and
   University of Massachusetts Amherst, USA
 }

\titlerunning{Getting More out of Large Language Models for Proofs}

\begin{document}

\maketitle

\begin{abstract}

Large language models have the potential to simplify formal theorem proving and make it more accessible. But how to get
the most out of these models is still an open question.
To answer this question, we take a step back and explore the
failure cases of these models using common prompting-based techniques.
Our talk will discuss these failure cases and what they can teach us
about how to get more out of these models.
\end{abstract}



%
%

\paragraph{Introduction}
\label{sect:introduction}

Formal theorem proving is a crucial but challenging task---and 
a natural fit for automation. Historically, most work on formal proof
automation has relied on symbolic techniques~\cite{ringer2019qed}, sometimes combined with neural tools \cite{deephol,gptf,wu2021tacticzero,lample2022hypertree,tactok,diva}. Recent advances in large language models
have improved their instruction-following and in-context learning capabilities, making it possible to build powerful proof
automation that elides symbolic search procedures altogether,
as shown by Baldur \cite{first2023baldur}.

How can we get more out of these advances in large language models?
To answer that, we examine the capabilities of the state-of-the-art language models GPT-3.5 Turbo and GPT-4 to prove theorems in Coq using common prompting-based
techniques. In particular, we conduct a fine-grained
analysis of model outputs on an example project.
Our emphasis is on the \textit{failure} cases---how these outputs commonly go wrong,
and what that can teach us about how to get more out of these models.\\

\begin{adjustwidth}{-2.5 cm}{-2.5 cm}\centering\begin{threeparttable}[!htb]
\scriptsize
\begin{tabular}{lrrrrrrrrrr}\toprule
&\multicolumn{3}{c}{\textbf{GPT-3.5-Turbo}} &\multicolumn{5}{c}{\textbf{GPT-4}} &\textbf{Proverbot} \\\cmidrule{2-10}
&FS-rand &FS-sim &ZS &FS-rand &FS-sim &ZS &FS+Lem &ZS+Lem &- \\\midrule
\textbf{\#Correct Proof} &10 &8 &0 &7 &8 &0 &14 &9 &- \\
\textbf{\#Proven Theorems} &6 &4 &0 &6 &7 &0 &7&5 &23 \\
\bottomrule
\end{tabular}
\caption{Results. 
\textbf{`FS'} stands for `few-shot', \textbf{`ZS'} stands for `zero-shot'.  \textbf{`+lemma' '}denotes providing  lemmas preceding the query theorem in the file in the context.}\label{tab: tab1}
\end{threeparttable}\end{adjustwidth}
\begin{figure}
\begin{minipage}{.37\columnwidth}
\centering
\begin{subfigure}{\columnwidth}\centering
    \includegraphics[width=\columnwidth]{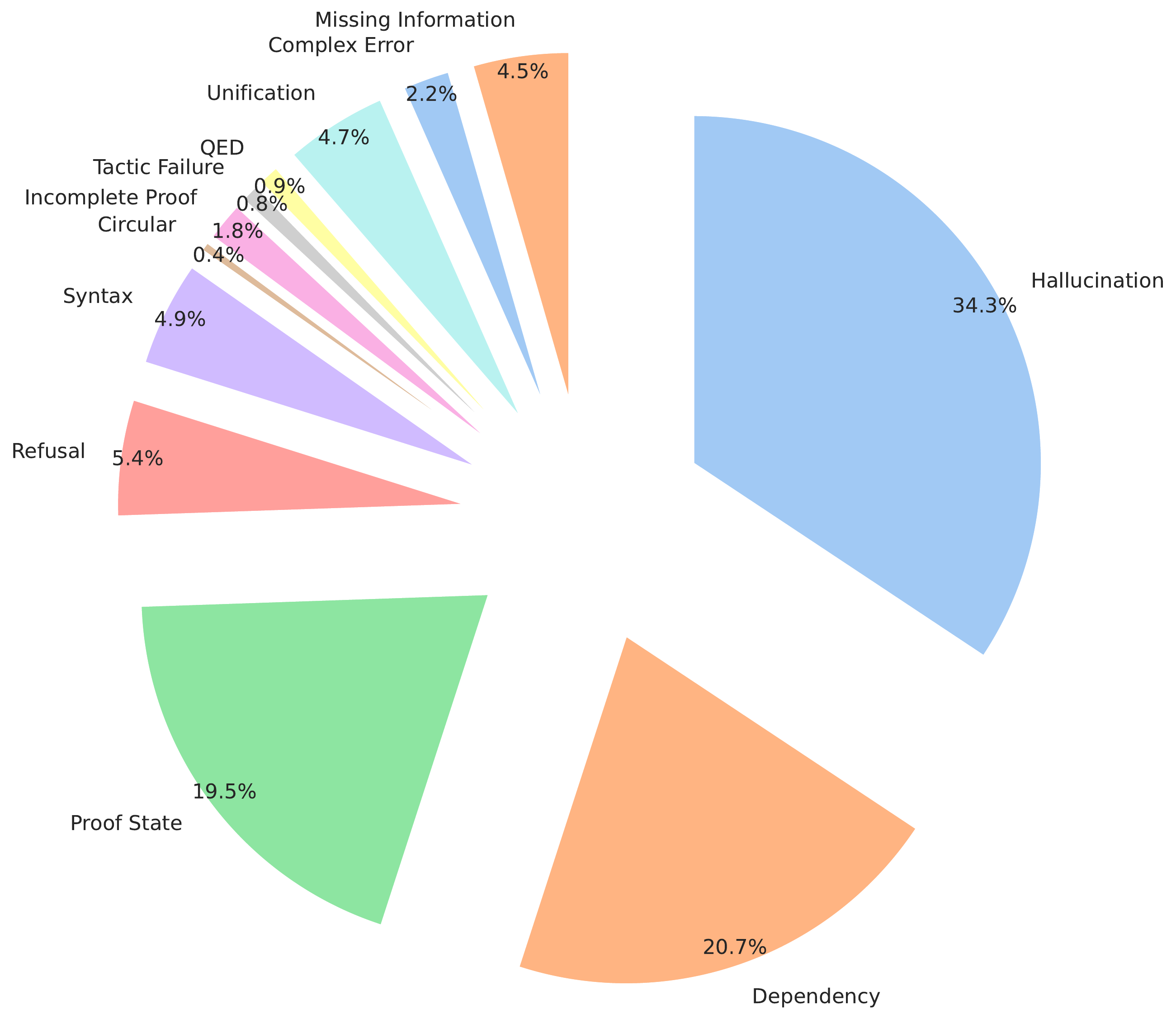}
    \caption{Breakdown of 740 expert-annotated GPT-4 outputs under few-shot `theorem-proof' set-up (see Appendix~\ref{theoremproof} for more).}
    \label{fig:breakdown}
\end{subfigure}
\end{minipage}
\begin{minipage}{0.6\columnwidth}
\hfill
\begin{subfigure}{.45\columnwidth}
       \centering
    \includegraphics[width=\columnwidth]{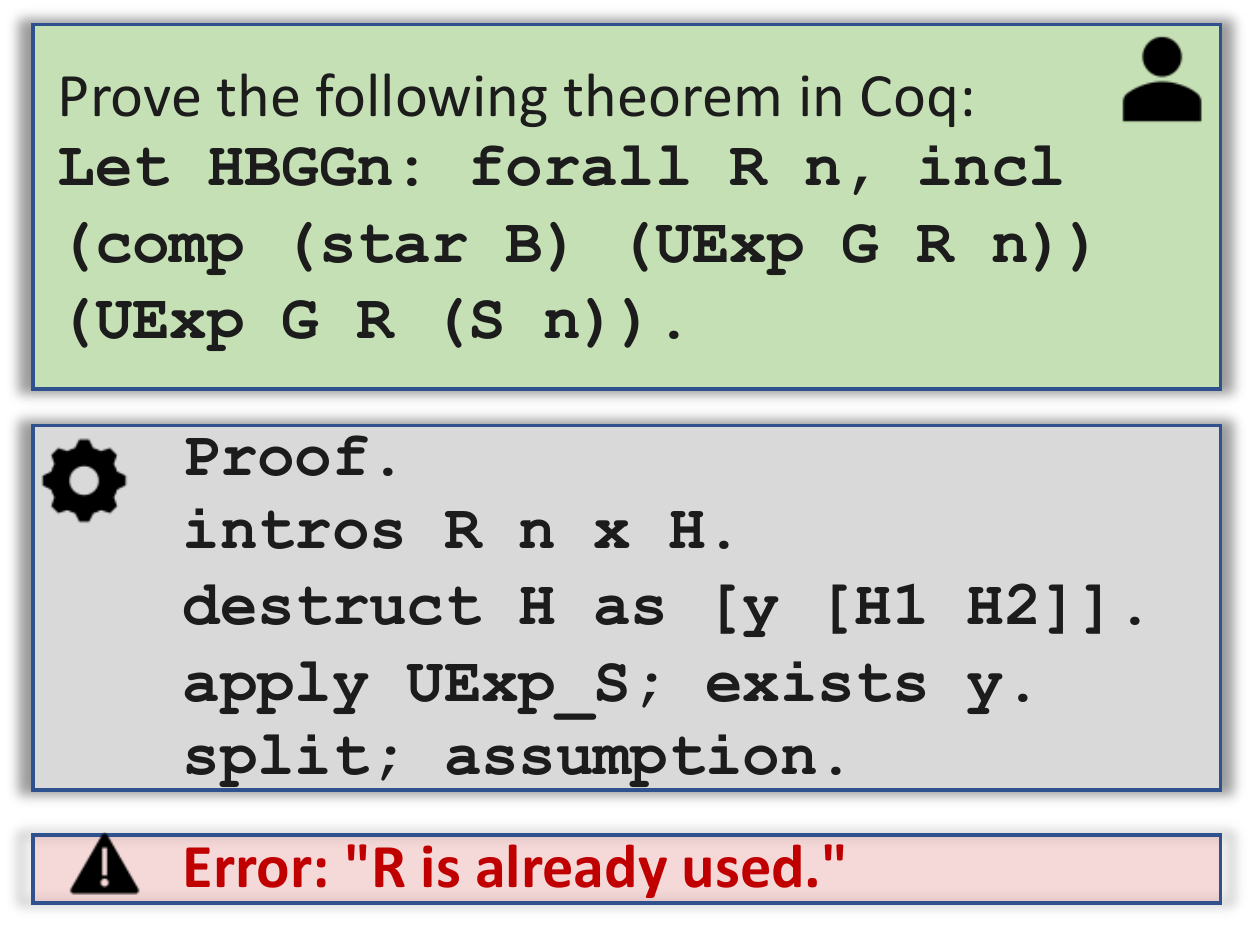}
    \caption{An example output that incurs an error because model lacks local proof state, and so uses variable name $R$ that is already taken. See Appendix~\ref{proofstate} for full context.}
    \label{fig:proofstate}
\end{subfigure}
\hfill
\begin{subfigure}{.45\columnwidth}\centering
    \includegraphics[width=\columnwidth]{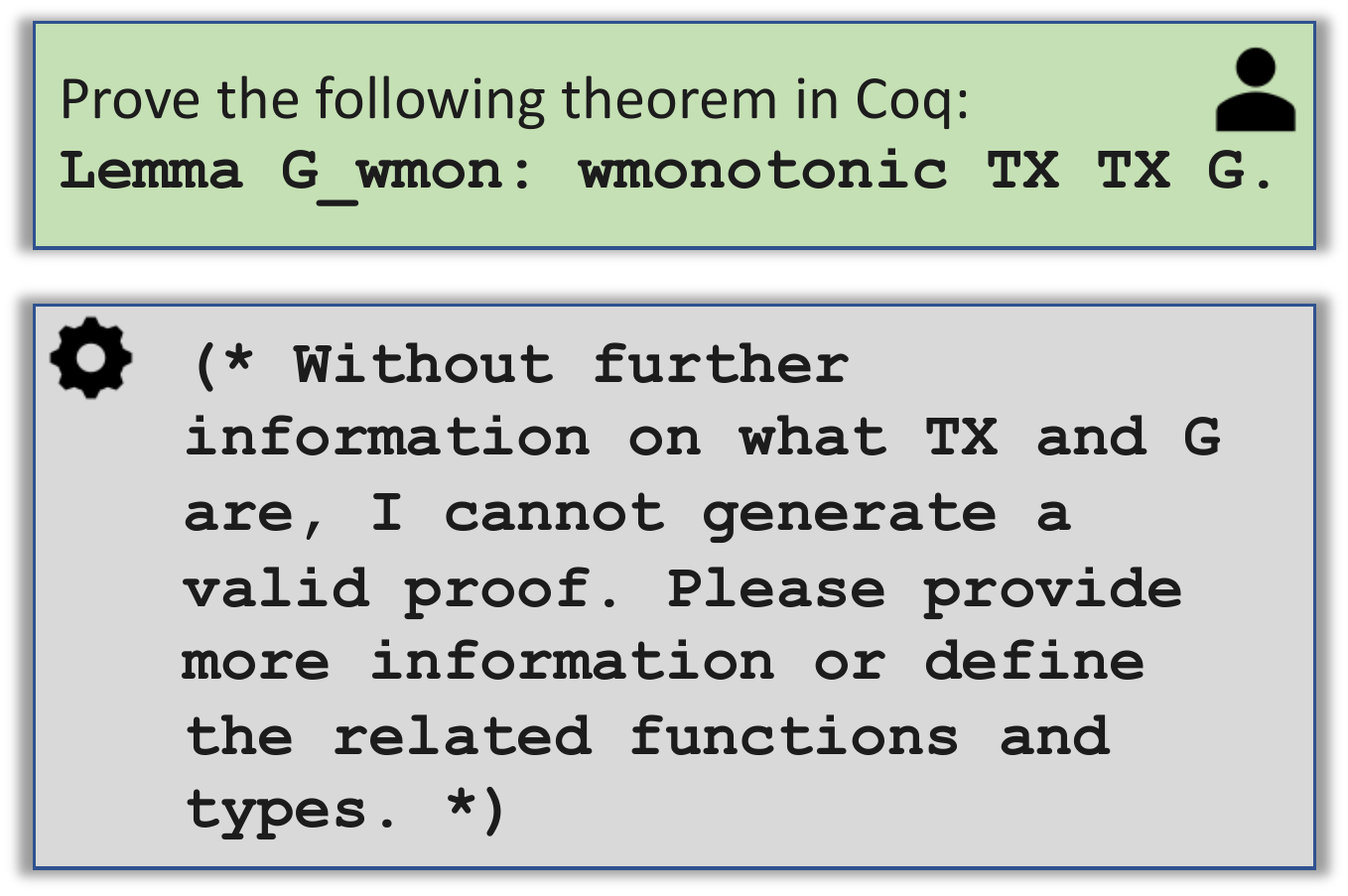}
    \caption{An example output that recognizes the necessary definitions are not provided and asks for clarifications from the user. See Appendix~\ref{sec:refusal} for more.}
    
    \label{fig:refuse}
\end{subfigure}
\end{minipage}
\end{figure}
\paragraph{Results and Recommendations}
We performed fine-grained analysis of the model-generated proofs and categorized the comments of human experts in Figure~\ref{fig:breakdown}. Details about our experimental methodology, along with examples, are in the appendix. Our recommendations:
\\\\
\textbf{\textit{1. Allow the model to prompt the proof assistant for more information.}}  Model outputs sometimes ask for more information about the definitions of referenced variables when those definitions have not been provided as input as shown in Figure~\ref{fig:refuse} and Appendix~\ref{sec:refusal}. This opens up a perfect opportunity for tool use at the model's request. By prompting the model to ask for information using standard Coq commands like \lstinline{Print}, and executing them in real time, we can allow the model to obtain information as it generates proof in steps.
\\\\
\textbf{\textit{2. Give the model access to proof states.} }Language models are found to be weak at `execution'. In our case, they by default lack access to proof state. This manifests in outputs as incorrect assumptions about the current proof state, like introducing too many variables or overloading variables already used (see Figure~\ref{fig:proofstate} and Appendix~\ref{proofstate}). For human proof engineers, interactive proof is more like a conversation with the proof assistant with constant feedback on the state;
    emulating this is a perfect opportunity to put
    the chat API to good use. 
\\\\
\textbf{\textit{3. Give the model access to information in file dependencies.}} Naively prompted models produce outputs that
    make incorrect assumptions about definitions and lemmas from dependencies,
    for example by hallucinating lemma names (see Appendix~\ref{sec:hallucination}), or by attempting to induct over
    a non-inductive hypothesis. Human proof engineers avoid this by having
    access to dependencies directly; models should access to those dependencies in context or in memory. 
\\\\
\textbf{\textit{4. Give the model access to proofs preceding the current proof.}} Naively prompted models fail to solve proofs that would be obvious from context, since human-written proofs are often small permutations of earlier proofs. Baldur~\cite{first2023baldur} showed in-context learning can help with this in Isabelle/HOL;
    we now have evidence (see Appendix~\ref{sec:preceding}) this is worth trying in Coq.
\\\\
\textbf{\textit{5. Learn from errors.}} People have attempted to iteratively improve the quality of generated programs by providing the model with correctness signals as feedback~\cite{xia2023conversational,wang2022compilable}; the same has shown success for proofs in Isabelle/HOL~\cite{first2023baldur}.
The error messages we observed were very informative,\footnote{Our annotated data with errors is available here: \url{https://github.com/DylanZSZ/LLM4Proof.git}} and suggest
this may be fruitful in Coq as well.
\\\\
\textbf{\textit{6. Introduce diversity through prompt engineering.}} It has been observed that a mixture of experts can boost performance of ML-guided proof synthesis by performing diverse sequences of tactics~\cite{diva}. On the other hand, there is evidence of performance gain by introducing diversity in few-shot set-up for both synthesis~\cite{austin2021program} and reasoning~\cite{wang2023selfconsistency}  tasks from the language modeling research community. The diversity we have already observed (see Appendix~\ref{sec:coinciding}) is evidence this may be worth trying with different prompts for Coq proofs.

\paragraph{Proposed Talk} In our talk, we will discuss specific examples of these failure cases, how they correspond to our recommendations, and our progress on experimentation along these recommendations in the intervening months.

\label{sect:bib}

\bibliographystyle{plain}
\bibliography{easychair}
\appendix
\section{Methodology}
We experimented with both zero-shot and few-shot prompting methods with \& without preceding lemmas of the query theorem. To select few-shot samples, we experimented with two strategies: random selection and similarity based selection using a retrieval model we fine-tuned on the training set. The details are deferred to appendix. 
\label{sect:typesetting}
\subsection{Zero-Shot Prompting}
The notable characteristic of GPT-3.5 Turbo and GPT-4 is their robust ability to follow instructions. In a zero-shot configuration, we leverage the system's messaging to establish the task and desired output format for the entire conversation, while utilizing user messaging to provide query-specific information such as the theorem to be proven and relevant lemmas.

\begin{figure}[htbp]
    \centering
    \begin{minipage}[t]{0.65\textwidth}
        \centering
        \includegraphics[width=\textwidth]{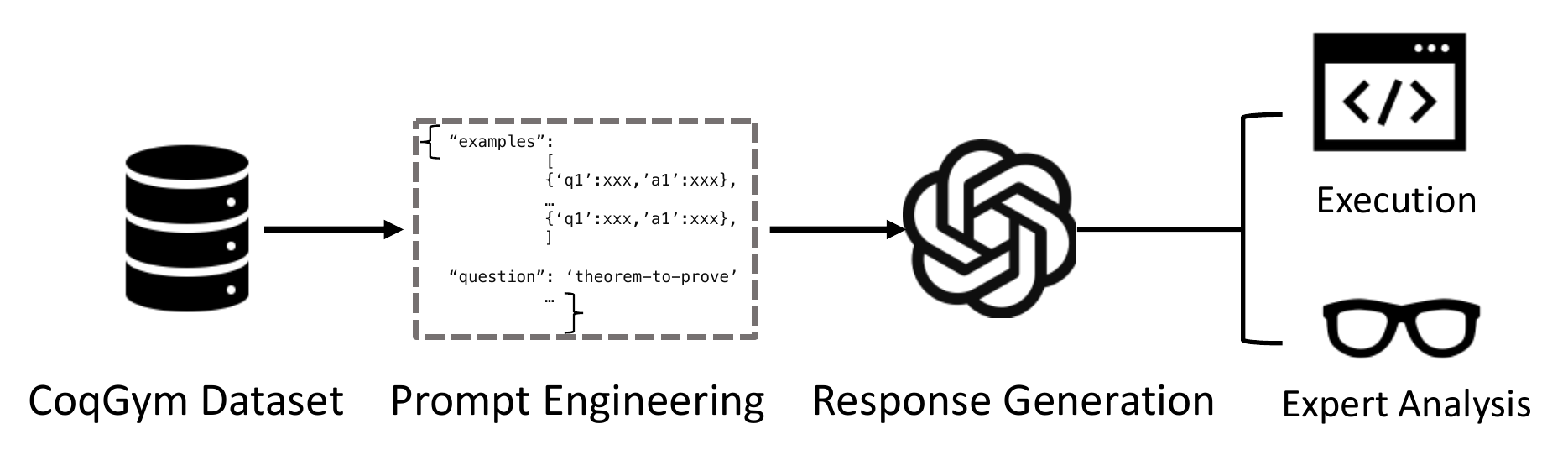}
        \caption{Investigation Pipeline}
        \label{fig:figure1}
    \end{minipage}
    \begin{minipage}[t]{0.25\textwidth}
        \centering
        \includegraphics[width=\textwidth]{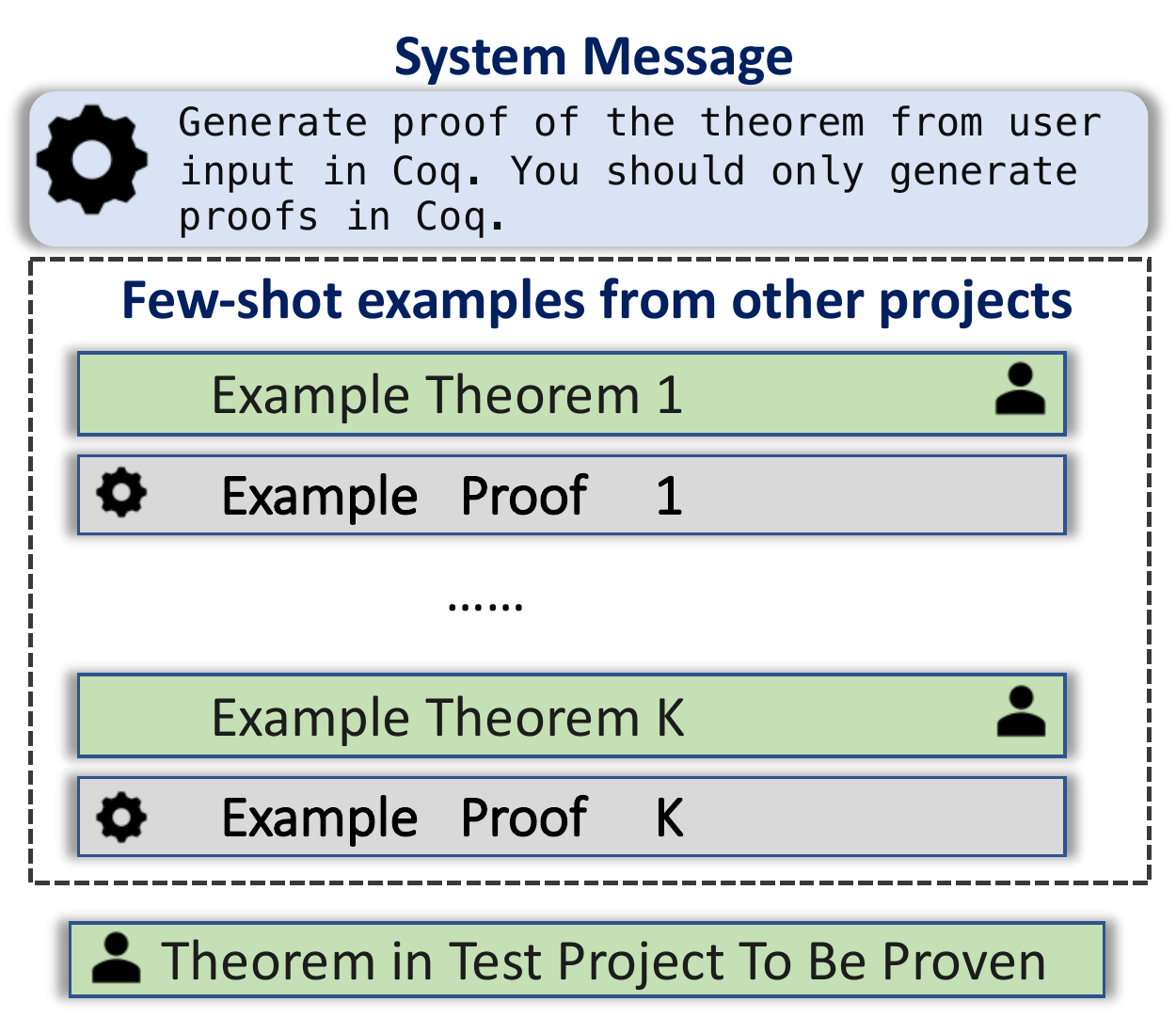}
        \caption{Workflow of few-shot prompting}
        \label{fig:few-shot}
    \end{minipage}
\end{figure}

\begin{figure}[htbp]
    \centering
    \begin{minipage}[t]{0.44\textwidth}
        \centering
        \includegraphics[width=\textwidth]{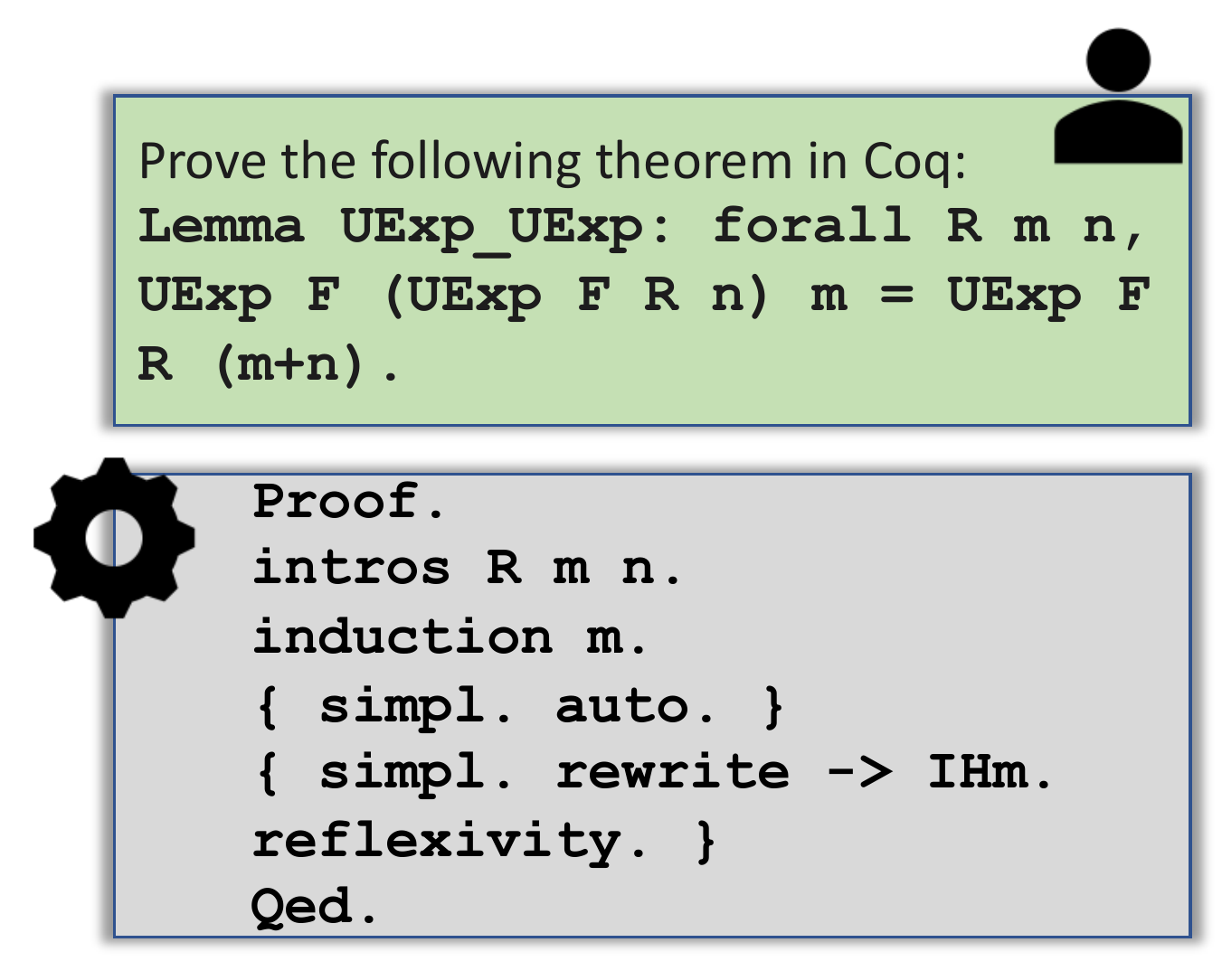}
        \caption{Few-shot example format of Theorem-Proof setting.}
        \label{fig:theoremproof}
    \end{minipage}
    \hfill
    \begin{minipage}[t]{0.44\textwidth}
        \centering
        \includegraphics[width=\textwidth]{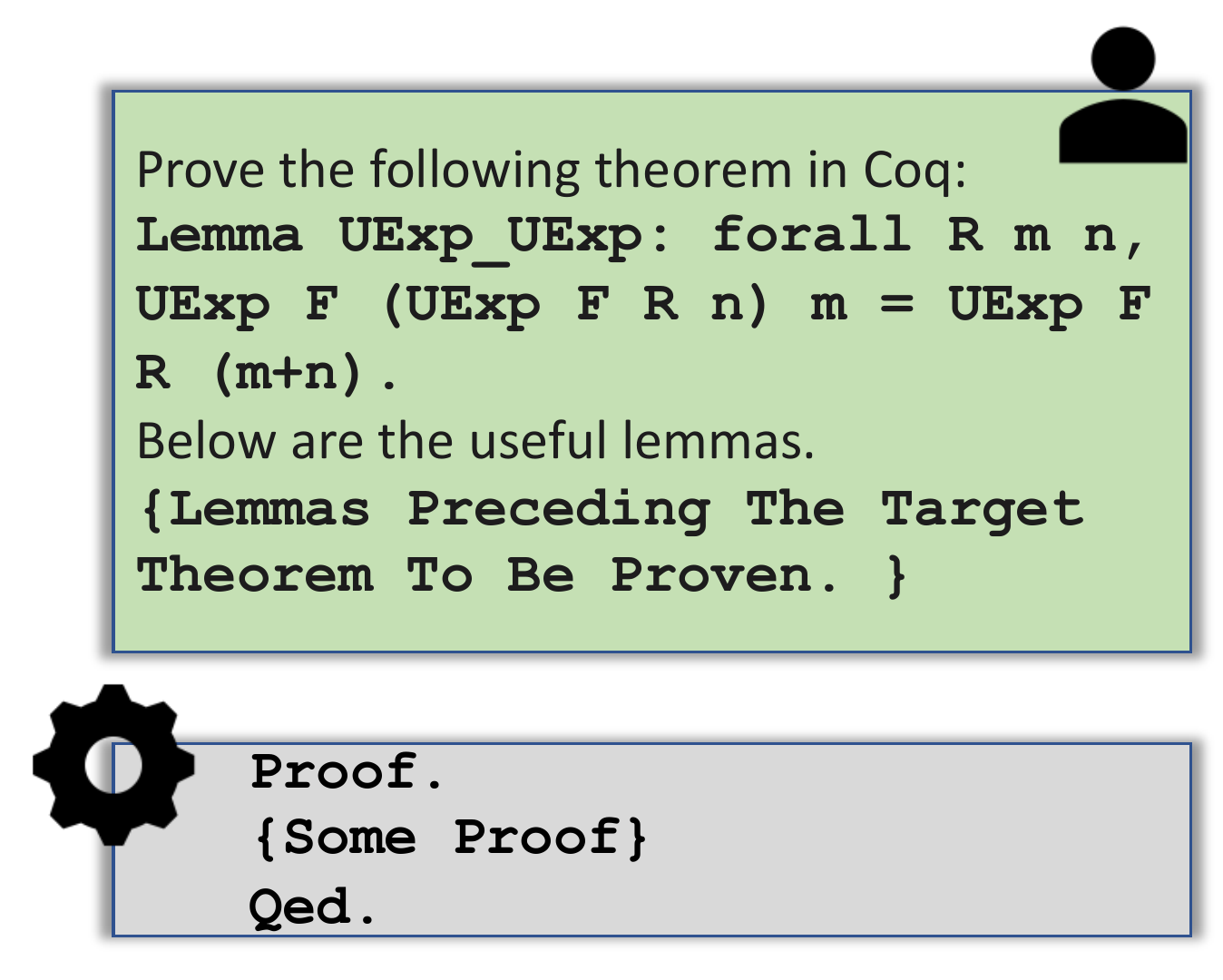}
        \caption{Few-shot example format of Theorem with Given Lemma setting.}
        \label{fig:givenlemma}
    \end{minipage}
\end{figure}
\subsection{Few-Shot Prompting}
Our approach utilizes the conventional few-shot in-context learning method. We include examples in the form of theorem-proof tuples at the beginning of the target query as shown in Figure~\ref{fig:few-shot}. In addition to the description of the task in the initial system message and the query, we provide examples of query-answer pairs obtained from training set to demonstrate to the model the task we want it to perform and the expected output format. 
We experimented with a few variants:
\paragraph{Theorem-Proof}\label{theoremproof}
In this set-up, we provide examples of pairs of theorems and proofs, as illustrated in Figure~\ref{fig:theoremproof}.
\paragraph{Theorem with Given Lemmas}
In this setting, we provide the model with the names of lemmas preceding the theorem. The format is shown in Figure~\ref{fig:givenlemma}.
\subsection{Training of Retriever}
We perform continued fine-tuning from the sentence-transformer model \textbf{all-mpnet-base-v2} based on\cite{song2020mpnet}. In particular, this model has been trained on code corpus and has certain level of code-understanding capabilities. We train the model to perform similar proof search on the train set by minimizing the triplet objective \begin{equation}
    J = \sum{_{i=1}^m} L_{triplet}(T_i, P_i, P_{j\ne i})
\end{equation}
where $T_i$ is the query theorem,$P_i$ is the proof of this theorem and $P_j$ is a randomly sampled proof of another theorem. During retrieval, we compute the similarity scores between test theorems and proofs in the training set, ans use those theorems whose proofs are similar to the test instance as the few-shot examples.

\subsection{Experiment Set-up}
We conducted experiments to evaluate the performance of two language models, GPT-3.5-turbo and GPT4, under both few-shot and zero-shot settings. The few-shot learning was performed with k-shots=6, meaning that the models were trained on only six examples per class.

The hyperparameters used for both models were set as follows: temperature $T$=1,presence
\_penalty=0.1, number of samples per prompt $n$=5. $presence\_penalty$ is a penalty term applied to the logit values of tokens already present in the prompt to discourage the repetition of those tokens in the generated text. 
\section{Coinciding proofs solved}
\label{sec:coinciding}

\begin{adjustwidth}{-2.5 cm}{-2.5 cm}\centering\begin{threeparttable}[!htb]
\caption{Number of coinciding theorems proven successfully by each pair of prompting set-ups.}\label{tab:coincide}
\scriptsize
\begin{tabular}{lrrrrrrr}\toprule
&\textbf{GPT4-random-fs} &\textbf{GPT3.5-random-fs} &\textbf{GPT4-sim-fs} &\textbf{GPT3.5-sim-fs} &\textbf{GPT4-lemma-fs} &\textbf{GPT4-lemma-zs} \\\midrule
\textbf{GPT4-random-fs} &- &- &- & &- &- \\
\textbf{GPT3.5-random-fs} &1 &- &- &- &- &- \\
\textbf{GPT4-sim-fs} &1 &2 &- &- &- &- \\
\textbf{GPT3.5-sim-fs} &0 &3 &1 &- &- &- \\
\textbf{GPT4-lemma-fs} &2 &1 &1 &0 &- &- \\
\textbf{GPT4-lemma-zs} &1 &1 &1 &0 &3 &- \\
\bottomrule
\end{tabular}
\end{threeparttable}\end{adjustwidth}
Proof synthesis is a special task in nature, having similarities with code-synthesis where both being generative, but still different as the criteria of whether or not the target theorem is proved is close-ended to a certain extent.
\section{Example Outputs}
\subsection{Refusing To Answer}
\label{sec:refusal}
\begin{lstlisting}
  Lemma G_wmon: wmonotonic TX TX G.
    (* Without further information on what TX and G are, I cannot generate a valid proof. Please provide more information or define the related functions and types. *)
\end{lstlisting}
Even under few-shot set-up where the model output is restricted to be in Coq format, the model insists on requesting additional information on the variables not clearly defined. It is impressive that the model has certain level of ability to understand Coq as well as the necessary information to complete the proof, e.g. the definition of variable names ans assumptions. This category constitutes 5.4\% of total number of cases.

\subsection{Not Trying Simple Tactics}
\begin{lstlisting}
  Lemma trans_incl: incl R R' -> incl (trans R) (trans R').
\end{lstlisting}
This is an easy theorem which model has failed. In particular, to prove this, a simple auto. tactic will do. However, the model did not give this tactic a try. 
\begin{lstlisting}
  Lemma trans_incl: incl R R' -> incl (trans R) (trans R').
  Proof.
    auto.
  Qed.
\end{lstlisting}
This is a case where all of context, proof state, and dependency information
may help. It is also a case where diverse prompting may be useful (for example, by adding a prompt that favors very simple proofs that do not use any lemmas).

\subsection{Succeeding}
There are certain cases where the model succeeded in proving the theorems. These theorems generally do not depend much on context. For example,
the model-generated proof of \lstinline{weak_refl} succeeds using only simple tactics:
\begin{lstlisting}
  Lemma weak_refl: forall x, Weak T x x.
  Proof.
    intros x.
    constructor.
    reflexivity.
  Qed.
\end{lstlisting}

\subsection{Lacking the Proof State}
\label{proofstate}
Here we provide the proof state of the example in the main text before the step in Figure~\ref{fig:proofstate}:
\begin{lstlisting}
  A, X, Y : Type
  TX : reduction_t A X
  TY : reduction_t A Y
  B : relation X
  HB : controlled TX TY B
  F, G : function X Y
  HF : monotonic TX TY F
  HG : wmonotonic TX TY G
  HBF : transparent B F
  HFG : contains F G
  HBG : contains (chaining_l (star B)) G
  R : relation2 X Y
  HRt : evolve_t TX TY R (comp (star B) (F R))
  HRa : evolve_a TX TY R (G R)
  pre_silent :
    forall n : nat,
      evolve_t TX TY (UExp F R n) (comp (star B) (UExp F R (S n)))
  silent : simulation_t TX TY (comp (star B) (UIter F R))
  HFGn : forall n : nat, incl (UExp F R n) (UExp G R n)
  ______________________________________(1/1)
  forall (R0 : relation2 X Y) (n : nat),
    incl (comp (star B) (UExp G R0 n)) (UExp G R0 (S n))
\end{lstlisting}
Note the proof assistant renamed the \lstinline{R} in the goal into \lstinline{R0} because there is another \lstinline{R} in context from earlier in the file.\footnote{Coq makes it possible to generalize all proofs in a \lstinline{Section} by the same pre-defined parameters, and one such parameter here is named \lstinline{R}.} Coq chooses \lstinline{R0}
so as not to shadow \lstinline{R}. Similarly,
the human proof calls this \lstinline{RR} to avoid shadowing \lstinline{R}:

\begin{lstlisting}
    intros RR n x y H; right; apply (HBG H).
\end{lstlisting} 
In contrast, without any access to the fact that \lstinline{R} is already defined the local proof state, the model output attempts to introduce a new
variable named \lstinline{R}. Coq refuses and responds with an error.

Another place where proof state is useful is when the model introduces
more variables than can be introduced. For example, the model
generates this output:

\begin{lstlisting}
  Lemma G_reverse: forall R, eeq (trans (G R)) (G (trans R)).
  Proof.
    unfold G, eeq.
    intros R u v.
    destruct R as [R Hr].
    simpl.
    split; intros [r H]; cbn in *; exists r;
    rewrite <- Hr in *;
    auto using sym_equal, trans_sym with relations.
\end{lstlisting}
The second step, \lstinline{intros R u v}, assumes there are three variables
to introduce. This would be reasonable if unfolding \lstinline{G} and \lstinline{eeq} introduced more \lstinline{forall}s in the goal of the proof state after the first step. 
But it does not; the local proof state at that step includes only one
variable \lstinline{R} to introduce. Coq thus responds with an error.

\subsection{Lacking Local Context}
\label{sec:preceding}

Writing proofs is often easier with context, since proofs often mirror preceding proofs in the same file. For example, proving this lemma is easier with context:

\begin{lstlisting}
  Lemma union_incl: (forall i, incl (F i) (F' i)) -> incl (union F) (union F').
\end{lstlisting}
This is because the human-written proofs preceding it in the file very much mirror the human-written proof of this lemma:

\begin{lstlisting}
  Lemma comp_incl: incl R R' -> incl S S' -> incl (comp R S) (comp R' S').
  Proof.
    unfold eeq, comp, incl; intuition.
    destruct H1 as [ t ]; exists t; auto.
  Qed.

  Lemma comp_eeq: eeq R R' -> eeq S S' -> eeq (comp R S) (comp R' S').
  Proof. 
    unfold eeq, comp, incl; intuition;
    destruct H0 as [ t ]; exists t; auto.
  Qed.

  Lemma union_incl: (forall i, incl (F i) (F' i)) -> incl (union F) (union F').
  Proof. 
    unfold eeq, union, incl; intuition.
    destruct H0 as [ i ]; exists i; auto. 
  Qed.
\end{lstlisting}
The same holds for the proof of \lstinline{union2_evolve_right}:

\begin{lstlisting}
  Lemma union2_evolve_left:
    forall l R S S', evolve_1 l R S -> evolve_1 l R (union2 S S').
  Proof.
    intros l R S S' H x x' y Hxx' xRy; destruct (H _ _ _ Hxx' xRy) as [ y' ]; 
    exists y'; auto; left; auto.
  Qed.
  
  Lemma union2_evolve_right:
    forall l R S S', evolve_1 l R S' -> evolve_1 l R (union2 S S').
  Proof.
    intros l R S S' H x x' y Hxx' xRy; destruct (H _ _ _ Hxx' xRy) as [ y' ]; 
    exists y'; auto; right; auto.
  Qed.
\end{lstlisting}
This gives us evidence that in-context learning in the style of Baldur may be fruitful in Coq as well.

\subsection{Hallucination}
\label{sec:hallucination}

Below is an example from the model where it hallucinates $H$ which has been defined nowhere. We observe in many cases the model are directly using the commonly used variable names like $H$, $x$, $y$ which it has seen multiple times during pre-training. 
\begin{lstlisting}
  Lemma bisimulation_bisim: bisimulation bisim.
  Proof.
    constructor.
    - intros.
      destruct H as [s' H].
      exists s'.
      apply stutter_bisim in H.
      auto.
\end{lstlisting}
Moreover, it hallucinates lemmas or definitions in the proofs, like \lstinline{stutter_bisim}. By providing proof state, we can help the model pick more correct variables; by providing file dependencies that contain referenced definitions and auxiliary lemmas, and prompting the model to restrict itself to those, we can help the model use only the definitions and lemmas that exist already.
\end{document}